\documentstyle{article}
\begin{document}
\title{Critical behavior of Josephson-junction arrays at $f=1/2$}
\author{Enzo Granato \\
International Centre for Theoretical Physics, 34100 Trieste, Italy \\
and Instituto Nacional de Pesquisas Espaciais, \\
12225 S\~ao Jos\'e dos
Campos, SP, Brazil \\
J.M. Kosterlitz \\
Department of Physics, \\
Brown University, Providence RI 02912, USA\\
M.P. Nightingale \\
Department of Physics, \\
University of Rhode Island, Kingston RI 02881, USA }

\date{\, }
\maketitle

\begin{abstract}
\hyphenation{pla-quet-te}
The critical behavior of frustrated Josephson-junction arrays at
$f=1/2$ flux quantum per plaquette is considered. Results from Monte
Carlo simulations and transfer matrix computations support the
identification of the critical behavior of the square and triangular
classical arrays and the one-dimensional quantum ladder with the
universality class of the $XY$-Ising model. In the quantum ladder, the
transition can happen either as a simultaneous ordering of the $Z_2$
and $U(1)$ order parameters or in two separate stages, depending on the
ratio between interchain and intrachain Josephson couplings. For the
classical arrays, weak random plaquette disorder acts like a random
field and positional disorder as random bonds on the $Z_2$ variables.
Increasing positional disorder  decouples the $Z_2$ and $U(1)$
variables leading to the same critical behavior as for integer $f$.
\end{abstract}

\newpage
\section{Introduction}
For an ordered two-dimensional array of Josephson junctions at rational
values of the flux quantum per plaquette $f = \Phi/ \Phi_o$, where
$\Phi$ is the magnetic flux through a plaquette and $\Phi_o$ the flux
quantum, the ground state is a pinned vortex lattice commensurate with
the underlying periodic pinning potential leading to discrete
symmetries in addition to the $U(1)$ symmetry of the superconducting
order parameter \cite{tj83}. When quantum fluctuations due to
charging  effects can be ignored,  yielding the so-called classical array, the
critical properties can be described by a frustrated $XY$
spin model in two dimensions with frustration parameter $f$.
Inclusion of charging effects adds an extra dimension, the time
direction, but one-dimensional arrays in the form of ladders can still
be described by two-dimensional classical spin models.  Of particular
interest is the case when $f = 1/2$ which has been intensively studied
both experimentally and theoretically because of the interplay between
continuous $U(1)$ and discrete $Z_2$ order parameters. The $Z_2$ Ising
symmetry can be associated with an antiferromagnetic arrangement of
plaquette chiralities, $\chi = \pm 1$, measuring the directions of the
circulating currents in each plaquette \cite{tj83,jv77,halsey85}.

It is quite natural to expect for a Josephson-junction array at
$f=1/2$, that critical behavior resulting from the interplay between
vortices and Ising degrees of freedom, can be described by some form of
coupled $XY$ and Ising models. Such models, containing both continuous
$U(1)$ and discrete $Z_2$ symmetries, can give rise to interesting and
sometimes unusual critical behavior \cite{gkln91}.  Recently, there has been
a lot of interest in these systems, not only in relation to
Josephson-junction arrays but also in surface phase transitions where
the interplay between roughening and reconstruction of the surface is
important \cite{mdn92,jug93,bast95}. A minimal model which is general
enough and consistent with the basic symmetries can be defined by the
$XY$-Ising model \cite{gkln91,eg87,lgk91}
\begin{equation}
\frac{H}{kT} = - \sum_{<ij>} [(A + B \sigma_i \sigma_j)
\cos(\theta_i-\theta_j)
 +C \sigma_i \sigma_j ]
\label{eq.ham}
\end{equation}
where $\sigma=\pm 1$ is an Ising spin and $\theta = [0,2\pi]$ is the
phase of a two-component unit vector ($XY$ spin). The Ising and $XY$
variables are coupled by their energy densities in the same way as two
independent systems of Ising spins are coupled in the Ashkin-Teller
model.  The model can also be regarded as the infinite coupling limit,
$ h \rightarrow \infty$, of two $XY$ models \cite{cd85,yd85,gk86}
coupled by a term of the form $-h \cos 2(\theta_1 - \theta_2)$.  In the
$A B$-plane, the model defined by Eq.~(\ref{eq.ham}) has a rich phase
diagram that depends strongly on the value of $C$. The model with $A
\ne B$ is relevant for an anisotropic frustrated $XY$ model
\cite{berge,gk86,eikmans} which can, in principle, be physically
realized as a modulated Josephson-junction array at $f=1/2$, where the
Josephson coupling in every other column of junctions is different by a
constant factor from the others.

For the isotropic square and triangular classical arrays \cite{lgk91} and for
the ladder quantum array \cite{eg92} the relevant subspace of model
(\ref{eq.ham}) is the symmetric one, $A=B$,
\begin{equation}
\frac{H}{kT} = - \sum_{<ij>} [A (1 +
\sigma_i \sigma_j) \cos(\theta_i-\theta_j) +C \sigma_i \sigma_j ].
\label{xyiAC}
\end{equation}
This case has a special symmetry since $XY$ variables are not coupled
across an Ising domain wall where $\sigma_i \sigma_j +1 = 0$.  In the
original frustrated $XY$  model, the phases across the chiral domain
walls are actually coupled in the ground state \cite{halsey85}. This
coupling and additional terms allowed by symmetry have always been
assumed to be irrelevant at criticality in the $XY$-Ising model
description. Recently, Knops et al \cite{knops94} have addressed this
question using transfer matrix calculations  and provided further
justification for the relation between the $XY$-Ising model and the
frustrated $XY$ model on a square lattice by showing that at the
transition the phase coupling across domain walls vanishes for
increasing system sizes.

The phase diagram of the $XY$-Ising model given by Eq.~(\ref{xyiAC}) as
inferred by Migdal-Kadanoff renormalization \cite{eg87} and Monte
Carlo simulations \cite{gkln91,lgk91} consists of three branches which
meet at multi-critical point $P$, as indicated in Fig.  \ref{phadiag}.
One of the branches, $PT$,  corresponds to  single transitions with
simultaneous loss of $XY$ and Ising order, and the other two to separate
Kosterlitz-Thouless and Ising transitions.  The line of single
transitions becomes first order at the tricritical point $T$.  The
critical behavior of a Josephson-junction array at $f=1/2$ as a
function of temperature corresponds to a particular cut through this
phase diagram and the single or double character of the transition
depends on the relative location of this path to the branch point $P$.
In Monte Carlo simulations \cite{gkln91}, the critical line $PT$ in the
phase diagram appeared to be non-universal as the critical exponents
associated with the $Z_2$ order parameter were found to vary
systematically along this line. More recent work based on transfer
matrix calculations \cite{ngk95} found no clear evidence for variation
of these exponents and suggested that the apparent non-universality may be
due to slow convergence and strong corrections to scaling.
However, it is clear that the values of the critical exponents
estimated by finite-size scaling of large systems are inconsistent with
pure Ising critical behavior, $\nu =1, \eta = 1/4$, as can be seen
from Table \ref{xyitab}.

\begin{table}[tbh]
\caption{Critical exponents $\nu$ and $\eta$ for the $Z_2$ order
parameter and central charge $c$ of the  $XY$-Ising model on the critical
line $PT$ (Fig. \protect \ref{phadiag}) at $A=1,\ C=-0.2885$. Results are
from Monte Carlo simulations (MC) and Monte Carlo transfer matrix
calculations (MCTM). The value of $c$ is the estimate for the largest
system without extrapolation}
\medskip
\begin{center}
\begin{tabular}{c|c|c|c|c}
\hline \hline
 \,     &\,                &$\nu$      &$\eta$      &$c$            \\
\hline
 MC     & Ref. \cite{lgk91}  &$0.84$     &$0.31$      &\,             \\
MCTM    & Ref. \cite{ngk95}  &$0.79$     &$0.40$      &$\sim 1.60$    \\
\hline \hline
\end{tabular}
\end{center}
\label{xyitab}
\end{table}

The central charge $c$, which provides additional information on the
nature of this critical line of simultaneous Ising and $XY$ transitions
was also evaluated by transfer matrix
calculations on infinite strips of widths as large as 30 lattice
spacings \cite{ngk95}. A value, $c = 1 + 1/2 $, would be expected if
the critical behavior is the result of a superposition of critical
Ising ($c=1/2$) and Gaussian ($c=1$) models \cite{foda}. The numerical
estimate for $c$ however is usually higher than this value, as indicated
in Table \ref{xyitab}, but effective values of $c$ obtained for
increasing system sizes decrease significantly.
Extrapolation assuming power-law corrections \cite{ngk95}
yield results not inconsistent with $c=3/2$.  These results suggest
that the critical line $PT$ is controlled by a non-trivial
fixed point with new critical exponents. In this case the phase transition in
the Josephson-junction array system at $f=1/2$ would be in a new
universality class, if this system indeed corresponds to a cut through the
$XY$-Ising model that intersects the critical line of simultaneous
$XY$ and Ising transitions.

\section{Classical arrays: the frustrated $XY$ model}
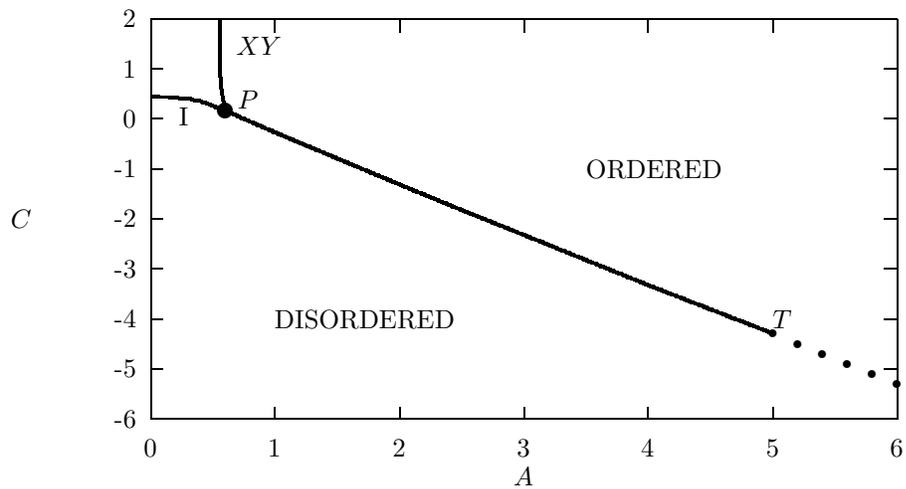
\begin{figure}[tbh]
\begin{center}

\setlength{\unitlength}{0.240900pt}
\ifx\plotpoint\undefined\newsavebox{\plotpoint}\fi
\sbox{\plotpoint}{\rule[-0.175pt]{0.350pt}{0.350pt}}%
\begin{picture}(1500,900)(0,0)
\tenrm
\sbox{\plotpoint}{\rule[-0.175pt]{0.350pt}{0.350pt}}%
\put(264,158){\rule[-0.175pt]{0.350pt}{151.526pt}}
\put(264,158){\rule[-0.175pt]{4.818pt}{0.350pt}}
\put(242,158){\makebox(0,0)[r]{-6}}
\put(1416,158){\rule[-0.175pt]{4.818pt}{0.350pt}}
\put(264,237){\rule[-0.175pt]{4.818pt}{0.350pt}}
\put(242,237){\makebox(0,0)[r]{-5}}
\put(1416,237){\rule[-0.175pt]{4.818pt}{0.350pt}}
\put(264,315){\rule[-0.175pt]{4.818pt}{0.350pt}}
\put(242,315){\makebox(0,0)[r]{-4}}
\put(1416,315){\rule[-0.175pt]{4.818pt}{0.350pt}}
\put(264,394){\rule[-0.175pt]{4.818pt}{0.350pt}}
\put(242,394){\makebox(0,0)[r]{-3}}
\put(1416,394){\rule[-0.175pt]{4.818pt}{0.350pt}}
\put(264,473){\rule[-0.175pt]{4.818pt}{0.350pt}}
\put(242,473){\makebox(0,0)[r]{-2}}
\put(1416,473){\rule[-0.175pt]{4.818pt}{0.350pt}}
\put(264,551){\rule[-0.175pt]{4.818pt}{0.350pt}}
\put(242,551){\makebox(0,0)[r]{-1}}
\put(1416,551){\rule[-0.175pt]{4.818pt}{0.350pt}}
\put(264,630){\rule[-0.175pt]{4.818pt}{0.350pt}}
\put(242,630){\makebox(0,0)[r]{0}}
\put(1416,630){\rule[-0.175pt]{4.818pt}{0.350pt}}
\put(264,708){\rule[-0.175pt]{4.818pt}{0.350pt}}
\put(242,708){\makebox(0,0)[r]{1}}
\put(1416,708){\rule[-0.175pt]{4.818pt}{0.350pt}}
\put(264,787){\rule[-0.175pt]{4.818pt}{0.350pt}}
\put(242,787){\makebox(0,0)[r]{2}}
\put(1416,787){\rule[-0.175pt]{4.818pt}{0.350pt}}
\put(264,158){\rule[-0.175pt]{0.350pt}{4.818pt}}
\put(264,113){\makebox(0,0){0}}
\put(264,767){\rule[-0.175pt]{0.350pt}{4.818pt}}
\put(459,158){\rule[-0.175pt]{0.350pt}{4.818pt}}
\put(459,113){\makebox(0,0){1}}
\put(459,767){\rule[-0.175pt]{0.350pt}{4.818pt}}
\put(655,158){\rule[-0.175pt]{0.350pt}{4.818pt}}
\put(655,113){\makebox(0,0){2}}
\put(655,767){\rule[-0.175pt]{0.350pt}{4.818pt}}
\put(850,158){\rule[-0.175pt]{0.350pt}{4.818pt}}
\put(850,113){\makebox(0,0){3}}
\put(850,767){\rule[-0.175pt]{0.350pt}{4.818pt}}
\put(1045,158){\rule[-0.175pt]{0.350pt}{4.818pt}}
\put(1045,113){\makebox(0,0){4}}
\put(1045,767){\rule[-0.175pt]{0.350pt}{4.818pt}}
\put(1241,158){\rule[-0.175pt]{0.350pt}{4.818pt}}
\put(1241,113){\makebox(0,0){5}}
\put(1241,767){\rule[-0.175pt]{0.350pt}{4.818pt}}
\put(1436,158){\rule[-0.175pt]{0.350pt}{4.818pt}}
\put(1436,113){\makebox(0,0){6}}
\put(1436,767){\rule[-0.175pt]{0.350pt}{4.818pt}}
\put(264,158){\rule[-0.175pt]{282.335pt}{0.350pt}}
\put(1436,158){\rule[-0.175pt]{0.350pt}{151.526pt}}
\put(264,787){\rule[-0.175pt]{282.335pt}{0.350pt}}
\put(45,472){\makebox(0,0)[l]{\shortstack{ $C$ }}}
\put(850,68){\makebox(0,0){ $A$ }}
\put(401,661){\makebox(0,0)[l]{$P$}}
\put(1241,315){\makebox(0,0)[l]{$T$}}
\put(309,634){\makebox(0,0)[l]{I}}
\put(399,744){\makebox(0,0)[l]{$XY$}}
\put(948,551){\makebox(0,0)[l]{ORDERED }}
\put(459,315){\makebox(0,0)[l]{DISORDERED}}
\put(264,158){\rule[-0.175pt]{0.350pt}{151.526pt}}
\sbox{\plotpoint}{\rule[-0.500pt]{1.000pt}{1.000pt}}%
\put(371,698){\rule[-0.500pt]{1.000pt}{21.380pt}}
\put(372,688){\rule[-0.500pt]{1.000pt}{2.349pt}}
\put(373,678){\rule[-0.500pt]{1.000pt}{2.349pt}}
\put(374,669){\rule[-0.500pt]{1.000pt}{2.349pt}}
\put(375,664){\rule[-0.500pt]{1.000pt}{1.084pt}}
\put(376,660){\rule[-0.500pt]{1.000pt}{1.084pt}}
\put(377,655){\rule[-0.500pt]{1.000pt}{1.084pt}}
\put(378,651){\rule[-0.500pt]{1.000pt}{1.084pt}}
\put(379,646){\rule[-0.500pt]{1.000pt}{1.084pt}}
\put(380,642){\rule[-0.500pt]{1.000pt}{1.084pt}}
\put(264,665){\usebox{\plotpoint}}
\put(264,665){\rule[-0.500pt]{3.553pt}{1.000pt}}
\put(278,664){\rule[-0.500pt]{3.553pt}{1.000pt}}
\put(293,663){\rule[-0.500pt]{3.553pt}{1.000pt}}
\put(308,662){\rule[-0.500pt]{3.553pt}{1.000pt}}
\put(323,661){\rule[-0.500pt]{1.144pt}{1.000pt}}
\put(327,660){\rule[-0.500pt]{1.144pt}{1.000pt}}
\put(332,659){\rule[-0.500pt]{1.144pt}{1.000pt}}
\put(337,658){\rule[-0.500pt]{1.144pt}{1.000pt}}
\put(342,657){\usebox{\plotpoint}}
\put(344,656){\usebox{\plotpoint}}
\put(347,655){\usebox{\plotpoint}}
\put(350,654){\usebox{\plotpoint}}
\put(353,653){\usebox{\plotpoint}}
\put(356,652){\usebox{\plotpoint}}
\put(359,651){\usebox{\plotpoint}}
\put(362,650){\usebox{\plotpoint}}
\put(363,649){\usebox{\plotpoint}}
\put(365,648){\usebox{\plotpoint}}
\put(367,647){\usebox{\plotpoint}}
\put(369,646){\usebox{\plotpoint}}
\put(370,645){\usebox{\plotpoint}}
\put(374,644){\usebox{\plotpoint}}
\put(377,643){\usebox{\plotpoint}}
\put(381,642){\usebox{\plotpoint}}
\put(383,641){\usebox{\plotpoint}}
\put(385,640){\usebox{\plotpoint}}
\put(387,639){\usebox{\plotpoint}}
\put(390,638){\usebox{\plotpoint}}
\put(392,637){\usebox{\plotpoint}}
\put(394,636){\usebox{\plotpoint}}
\put(397,635){\usebox{\plotpoint}}
\put(399,634){\usebox{\plotpoint}}
\put(401,633){\usebox{\plotpoint}}
\put(403,632){\usebox{\plotpoint}}
\put(406,631){\usebox{\plotpoint}}
\put(408,630){\usebox{\plotpoint}}
\put(410,629){\usebox{\plotpoint}}
\put(413,628){\usebox{\plotpoint}}
\put(415,627){\usebox{\plotpoint}}
\put(417,626){\usebox{\plotpoint}}
\put(420,625){\usebox{\plotpoint}}
\put(422,624){\usebox{\plotpoint}}
\put(424,623){\usebox{\plotpoint}}
\put(426,622){\usebox{\plotpoint}}
\put(429,621){\usebox{\plotpoint}}
\put(431,620){\usebox{\plotpoint}}
\put(433,619){\usebox{\plotpoint}}
\put(436,618){\usebox{\plotpoint}}
\put(438,617){\usebox{\plotpoint}}
\put(440,616){\usebox{\plotpoint}}
\put(442,615){\usebox{\plotpoint}}
\put(445,614){\usebox{\plotpoint}}
\put(447,613){\usebox{\plotpoint}}
\put(449,612){\usebox{\plotpoint}}
\put(452,611){\usebox{\plotpoint}}
\put(454,610){\usebox{\plotpoint}}
\put(456,609){\usebox{\plotpoint}}
\put(459,608){\usebox{\plotpoint}}
\put(461,607){\usebox{\plotpoint}}
\put(463,606){\usebox{\plotpoint}}
\put(466,605){\usebox{\plotpoint}}
\put(468,604){\usebox{\plotpoint}}
\put(470,603){\usebox{\plotpoint}}
\put(473,602){\usebox{\plotpoint}}
\put(475,601){\usebox{\plotpoint}}
\put(478,600){\usebox{\plotpoint}}
\put(480,599){\usebox{\plotpoint}}
\put(482,598){\usebox{\plotpoint}}
\put(485,597){\usebox{\plotpoint}}
\put(487,596){\usebox{\plotpoint}}
\put(490,595){\usebox{\plotpoint}}
\put(492,594){\usebox{\plotpoint}}
\put(494,593){\usebox{\plotpoint}}
\put(497,592){\usebox{\plotpoint}}
\put(499,591){\usebox{\plotpoint}}
\put(502,590){\usebox{\plotpoint}}
\put(504,589){\usebox{\plotpoint}}
\put(506,588){\usebox{\plotpoint}}
\put(509,587){\usebox{\plotpoint}}
\put(511,586){\usebox{\plotpoint}}
\put(513,585){\usebox{\plotpoint}}
\put(516,584){\usebox{\plotpoint}}
\put(518,583){\usebox{\plotpoint}}
\put(521,582){\usebox{\plotpoint}}
\put(523,581){\usebox{\plotpoint}}
\put(525,580){\usebox{\plotpoint}}
\put(528,579){\usebox{\plotpoint}}
\put(530,578){\usebox{\plotpoint}}
\put(533,577){\usebox{\plotpoint}}
\put(535,576){\usebox{\plotpoint}}
\put(537,575){\usebox{\plotpoint}}
\put(540,574){\usebox{\plotpoint}}
\put(542,573){\usebox{\plotpoint}}
\put(545,572){\usebox{\plotpoint}}
\put(547,571){\usebox{\plotpoint}}
\put(549,570){\usebox{\plotpoint}}
\put(552,569){\usebox{\plotpoint}}
\put(554,568){\usebox{\plotpoint}}
\put(557,567){\usebox{\plotpoint}}
\put(559,566){\usebox{\plotpoint}}
\put(561,565){\usebox{\plotpoint}}
\put(564,564){\usebox{\plotpoint}}
\put(566,563){\usebox{\plotpoint}}
\put(568,562){\usebox{\plotpoint}}
\put(571,561){\usebox{\plotpoint}}
\put(573,560){\usebox{\plotpoint}}
\put(576,559){\usebox{\plotpoint}}
\put(578,558){\usebox{\plotpoint}}
\put(580,557){\usebox{\plotpoint}}
\put(583,556){\usebox{\plotpoint}}
\put(585,555){\usebox{\plotpoint}}
\put(588,554){\usebox{\plotpoint}}
\put(590,553){\usebox{\plotpoint}}
\put(592,552){\usebox{\plotpoint}}
\put(595,551){\usebox{\plotpoint}}
\put(597,550){\usebox{\plotpoint}}
\put(600,549){\usebox{\plotpoint}}
\put(602,548){\usebox{\plotpoint}}
\put(604,547){\usebox{\plotpoint}}
\put(607,546){\usebox{\plotpoint}}
\put(609,545){\usebox{\plotpoint}}
\put(611,544){\usebox{\plotpoint}}
\put(614,543){\usebox{\plotpoint}}
\put(616,542){\usebox{\plotpoint}}
\put(619,541){\usebox{\plotpoint}}
\put(621,540){\usebox{\plotpoint}}
\put(623,539){\usebox{\plotpoint}}
\put(626,538){\usebox{\plotpoint}}
\put(628,537){\usebox{\plotpoint}}
\put(631,536){\usebox{\plotpoint}}
\put(633,535){\usebox{\plotpoint}}
\put(635,534){\usebox{\plotpoint}}
\put(638,533){\usebox{\plotpoint}}
\put(640,532){\usebox{\plotpoint}}
\put(643,531){\usebox{\plotpoint}}
\put(645,530){\usebox{\plotpoint}}
\put(647,529){\usebox{\plotpoint}}
\put(650,528){\usebox{\plotpoint}}
\put(652,527){\usebox{\plotpoint}}
\put(655,526){\usebox{\plotpoint}}
\put(657,525){\usebox{\plotpoint}}
\put(659,524){\usebox{\plotpoint}}
\put(662,523){\usebox{\plotpoint}}
\put(664,522){\usebox{\plotpoint}}
\put(667,521){\usebox{\plotpoint}}
\put(669,520){\usebox{\plotpoint}}
\put(671,519){\usebox{\plotpoint}}
\put(674,518){\usebox{\plotpoint}}
\put(676,517){\usebox{\plotpoint}}
\put(679,516){\usebox{\plotpoint}}
\put(681,515){\usebox{\plotpoint}}
\put(684,514){\usebox{\plotpoint}}
\put(686,513){\usebox{\plotpoint}}
\put(688,512){\usebox{\plotpoint}}
\put(691,511){\usebox{\plotpoint}}
\put(693,510){\usebox{\plotpoint}}
\put(696,509){\usebox{\plotpoint}}
\put(698,508){\usebox{\plotpoint}}
\put(701,507){\usebox{\plotpoint}}
\put(703,506){\usebox{\plotpoint}}
\put(705,505){\usebox{\plotpoint}}
\put(708,504){\usebox{\plotpoint}}
\put(710,503){\usebox{\plotpoint}}
\put(713,502){\usebox{\plotpoint}}
\put(715,501){\usebox{\plotpoint}}
\put(718,500){\usebox{\plotpoint}}
\put(720,499){\usebox{\plotpoint}}
\put(722,498){\usebox{\plotpoint}}
\put(725,497){\usebox{\plotpoint}}
\put(727,496){\usebox{\plotpoint}}
\put(730,495){\usebox{\plotpoint}}
\put(732,494){\usebox{\plotpoint}}
\put(735,493){\usebox{\plotpoint}}
\put(737,492){\usebox{\plotpoint}}
\put(739,491){\usebox{\plotpoint}}
\put(742,490){\usebox{\plotpoint}}
\put(744,489){\usebox{\plotpoint}}
\put(747,488){\usebox{\plotpoint}}
\put(749,487){\usebox{\plotpoint}}
\put(751,486){\usebox{\plotpoint}}
\put(754,485){\usebox{\plotpoint}}
\put(757,484){\usebox{\plotpoint}}
\put(759,483){\usebox{\plotpoint}}
\put(762,482){\usebox{\plotpoint}}
\put(764,481){\usebox{\plotpoint}}
\put(767,480){\usebox{\plotpoint}}
\put(769,479){\usebox{\plotpoint}}
\put(772,478){\usebox{\plotpoint}}
\put(774,477){\usebox{\plotpoint}}
\put(777,476){\usebox{\plotpoint}}
\put(779,475){\usebox{\plotpoint}}
\put(782,474){\usebox{\plotpoint}}
\put(784,473){\usebox{\plotpoint}}
\put(787,472){\usebox{\plotpoint}}
\put(789,471){\usebox{\plotpoint}}
\put(792,470){\usebox{\plotpoint}}
\put(794,469){\usebox{\plotpoint}}
\put(797,468){\usebox{\plotpoint}}
\put(799,467){\usebox{\plotpoint}}
\put(802,466){\usebox{\plotpoint}}
\put(804,465){\usebox{\plotpoint}}
\put(807,464){\usebox{\plotpoint}}
\put(809,463){\usebox{\plotpoint}}
\put(812,462){\usebox{\plotpoint}}
\put(814,461){\usebox{\plotpoint}}
\put(817,460){\usebox{\plotpoint}}
\put(819,459){\usebox{\plotpoint}}
\put(822,458){\usebox{\plotpoint}}
\put(824,457){\usebox{\plotpoint}}
\put(827,456){\usebox{\plotpoint}}
\put(829,455){\usebox{\plotpoint}}
\put(832,454){\usebox{\plotpoint}}
\put(834,453){\usebox{\plotpoint}}
\put(837,452){\usebox{\plotpoint}}
\put(839,451){\usebox{\plotpoint}}
\put(842,450){\usebox{\plotpoint}}
\put(844,449){\usebox{\plotpoint}}
\put(847,448){\usebox{\plotpoint}}
\put(849,447){\usebox{\plotpoint}}
\put(852,446){\usebox{\plotpoint}}
\put(854,445){\usebox{\plotpoint}}
\put(857,444){\usebox{\plotpoint}}
\put(859,443){\usebox{\plotpoint}}
\put(862,442){\usebox{\plotpoint}}
\put(864,441){\usebox{\plotpoint}}
\put(867,440){\usebox{\plotpoint}}
\put(869,439){\usebox{\plotpoint}}
\put(872,438){\usebox{\plotpoint}}
\put(874,437){\usebox{\plotpoint}}
\put(877,436){\usebox{\plotpoint}}
\put(879,435){\usebox{\plotpoint}}
\put(882,434){\usebox{\plotpoint}}
\put(884,433){\usebox{\plotpoint}}
\put(887,432){\usebox{\plotpoint}}
\put(889,431){\usebox{\plotpoint}}
\put(891,430){\usebox{\plotpoint}}
\put(894,429){\usebox{\plotpoint}}
\put(896,428){\usebox{\plotpoint}}
\put(899,427){\usebox{\plotpoint}}
\put(901,426){\usebox{\plotpoint}}
\put(904,425){\usebox{\plotpoint}}
\put(906,424){\usebox{\plotpoint}}
\put(909,423){\usebox{\plotpoint}}
\put(911,422){\usebox{\plotpoint}}
\put(914,421){\usebox{\plotpoint}}
\put(916,420){\usebox{\plotpoint}}
\put(919,419){\usebox{\plotpoint}}
\put(921,418){\usebox{\plotpoint}}
\put(924,417){\usebox{\plotpoint}}
\put(926,416){\usebox{\plotpoint}}
\put(928,415){\usebox{\plotpoint}}
\put(931,414){\usebox{\plotpoint}}
\put(933,413){\usebox{\plotpoint}}
\put(936,412){\usebox{\plotpoint}}
\put(938,411){\usebox{\plotpoint}}
\put(941,410){\usebox{\plotpoint}}
\put(943,409){\usebox{\plotpoint}}
\put(946,408){\usebox{\plotpoint}}
\put(948,407){\usebox{\plotpoint}}
\put(951,406){\usebox{\plotpoint}}
\put(953,405){\usebox{\plotpoint}}
\put(956,404){\usebox{\plotpoint}}
\put(958,403){\usebox{\plotpoint}}
\put(961,402){\usebox{\plotpoint}}
\put(963,401){\usebox{\plotpoint}}
\put(966,400){\usebox{\plotpoint}}
\put(968,399){\usebox{\plotpoint}}
\put(970,398){\usebox{\plotpoint}}
\put(973,397){\usebox{\plotpoint}}
\put(975,396){\usebox{\plotpoint}}
\put(978,395){\usebox{\plotpoint}}
\put(980,394){\usebox{\plotpoint}}
\put(983,393){\usebox{\plotpoint}}
\put(985,392){\usebox{\plotpoint}}
\put(988,391){\usebox{\plotpoint}}
\put(990,390){\usebox{\plotpoint}}
\put(993,389){\usebox{\plotpoint}}
\put(995,388){\usebox{\plotpoint}}
\put(998,387){\usebox{\plotpoint}}
\put(1000,386){\usebox{\plotpoint}}
\put(1003,385){\usebox{\plotpoint}}
\put(1005,384){\usebox{\plotpoint}}
\put(1007,383){\usebox{\plotpoint}}
\put(1010,382){\usebox{\plotpoint}}
\put(1012,381){\usebox{\plotpoint}}
\put(1015,380){\usebox{\plotpoint}}
\put(1017,379){\usebox{\plotpoint}}
\put(1020,378){\usebox{\plotpoint}}
\put(1022,377){\usebox{\plotpoint}}
\put(1025,376){\usebox{\plotpoint}}
\put(1027,375){\usebox{\plotpoint}}
\put(1030,374){\usebox{\plotpoint}}
\put(1032,373){\usebox{\plotpoint}}
\put(1035,372){\usebox{\plotpoint}}
\put(1037,371){\usebox{\plotpoint}}
\put(1040,370){\usebox{\plotpoint}}
\put(1042,369){\usebox{\plotpoint}}
\put(1045,368){\usebox{\plotpoint}}
\put(1047,367){\usebox{\plotpoint}}
\put(1050,366){\usebox{\plotpoint}}
\put(1052,365){\usebox{\plotpoint}}
\put(1055,364){\usebox{\plotpoint}}
\put(1057,363){\usebox{\plotpoint}}
\put(1060,362){\usebox{\plotpoint}}
\put(1063,361){\usebox{\plotpoint}}
\put(1065,360){\usebox{\plotpoint}}
\put(1068,359){\usebox{\plotpoint}}
\put(1070,358){\usebox{\plotpoint}}
\put(1073,357){\usebox{\plotpoint}}
\put(1075,356){\usebox{\plotpoint}}
\put(1078,355){\usebox{\plotpoint}}
\put(1081,354){\usebox{\plotpoint}}
\put(1083,353){\usebox{\plotpoint}}
\put(1086,352){\usebox{\plotpoint}}
\put(1088,351){\usebox{\plotpoint}}
\put(1091,350){\usebox{\plotpoint}}
\put(1094,349){\usebox{\plotpoint}}
\put(1096,348){\usebox{\plotpoint}}
\put(1099,347){\usebox{\plotpoint}}
\put(1101,346){\usebox{\plotpoint}}
\put(1104,345){\usebox{\plotpoint}}
\put(1106,344){\usebox{\plotpoint}}
\put(1109,343){\usebox{\plotpoint}}
\put(1112,342){\usebox{\plotpoint}}
\put(1114,341){\usebox{\plotpoint}}
\put(1117,340){\usebox{\plotpoint}}
\put(1119,339){\usebox{\plotpoint}}
\put(1122,338){\usebox{\plotpoint}}
\put(1124,337){\usebox{\plotpoint}}
\put(1127,336){\usebox{\plotpoint}}
\put(1130,335){\usebox{\plotpoint}}
\put(1132,334){\usebox{\plotpoint}}
\put(1135,333){\usebox{\plotpoint}}
\put(1137,332){\usebox{\plotpoint}}
\put(1140,331){\usebox{\plotpoint}}
\put(1143,330){\usebox{\plotpoint}}
\put(1145,329){\usebox{\plotpoint}}
\put(1148,328){\usebox{\plotpoint}}
\put(1150,327){\usebox{\plotpoint}}
\put(1153,326){\usebox{\plotpoint}}
\put(1155,325){\usebox{\plotpoint}}
\put(1158,324){\usebox{\plotpoint}}
\put(1161,323){\usebox{\plotpoint}}
\put(1163,322){\usebox{\plotpoint}}
\put(1166,321){\usebox{\plotpoint}}
\put(1168,320){\usebox{\plotpoint}}
\put(1171,319){\usebox{\plotpoint}}
\put(1173,318){\usebox{\plotpoint}}
\put(1176,317){\usebox{\plotpoint}}
\put(1179,316){\usebox{\plotpoint}}
\put(1181,315){\usebox{\plotpoint}}
\put(1184,314){\usebox{\plotpoint}}
\put(1186,313){\usebox{\plotpoint}}
\put(1189,312){\usebox{\plotpoint}}
\put(1192,311){\usebox{\plotpoint}}
\put(1194,310){\usebox{\plotpoint}}
\put(1197,309){\usebox{\plotpoint}}
\put(1199,308){\usebox{\plotpoint}}
\put(1202,307){\usebox{\plotpoint}}
\put(1204,306){\usebox{\plotpoint}}
\put(1207,305){\usebox{\plotpoint}}
\put(1210,304){\usebox{\plotpoint}}
\put(1212,303){\usebox{\plotpoint}}
\put(1215,302){\usebox{\plotpoint}}
\put(1217,301){\usebox{\plotpoint}}
\put(1220,300){\usebox{\plotpoint}}
\put(1222,299){\usebox{\plotpoint}}
\put(1225,298){\usebox{\plotpoint}}
\put(1228,297){\usebox{\plotpoint}}
\put(1230,296){\usebox{\plotpoint}}
\put(1233,295){\usebox{\plotpoint}}
\put(1235,294){\usebox{\plotpoint}}
\put(1238,293){\usebox{\plotpoint}}
\sbox{\plotpoint}{\rule[-0.175pt]{0.350pt}{0.350pt}}%
\put(1241,292){\circle*{12}}
\put(1280,276){\circle*{12}}
\put(1319,260){\circle*{12}}
\put(1358,244){\circle*{12}}
\put(1397,229){\circle*{12}}
\put(1436,213){\circle*{12}}
\put(381,642){\circle*{24}}
\end{picture}

\caption{ Phase diagram of the $XY$-Ising model obtained by Monte Carlo
simulations \protect{\cite{gkln91,lgk91}}.  Solid and dotted lines indicate
continuous and first-order transitions, respectively. The precise
locations of $P$ and $T$ are uncertain.}
\label{phadiag}
\end{center}
\end{figure}

There have been several numerical studies of the frustrated $XY$ model
on triangular and square lattices, but so far no definitive conclusion
has been reached as to the nature of the transition.  These are the
simplest models for a two-dimensional array at $f=1/2$ when capacitive
effects can be neglected \cite{tj83} and the critical behavior
describes the superconducting to normal transition in these systems.
Different numerical techniques, including finite-size analysis, seem to
lead to conflicting results which suggest either a single
\cite{tk90,lkg91,gn93,jose92,knops94} or double transitions
\cite{ms84,grest,lee94,olsson}.  It is clear however, that if there is
no single transition then there are two transitions which occur at
temperatures very close to each other and in a generalized version of
the  model they could join depending on the effective parameter
controlling the $XY$ and Ising excitations. In fact, a Coulomb-gas
representation of the frustrated $XY$ model with an additional coupling
between nearest-neighbor vortices suggests a mechanism that reconciles
the apparently contradictory results:  this model has a phase diagram
\cite{tk88} with similar structure as in Fig. \ref{phadiag} for the
$XY$-Ising model with the occurrence of double or single transitions
depending on this additional coupling.  Possibly a Josephson-junction
array at $f=1/2$ corresponds to a cut through the phase diagram of the
$XY$-Ising model close to the branch point $P$. If the critical
behavior associated with the $U(1)$ and $Z_2$ symmetries are determined
separately, the single or double nature of the transition is likely to
remain unresolved on purely numerical grounds. Computing the critical
point of a Kosterlitz-Thouless transition even in the simplest case is
problematic and requires scaling forms that lead to highly unstable
fits. As a consequence, the error bars of the critical point make
resolving two critical points impossible in many cases.

If the frustrated $XY$ models and the coupled $XY$-Ising model are in fact
in the same universality class then in order to verify the single
nature of the transition it is sufficient to study the $Z_2$ degrees of
freedom \cite{gkln91,lkg91}.  If the critical exponents are
inconsistent with pure Ising values, the transition cannot correspond
to the Ising branch of a double transition as  in Fig. \ref{phadiag}.
This point of view has the advantage of not requiring  a precise study
of the behavior of the $XY$ variables which for a model of this nature
cannot be interpreted using what is known from the standard $XY$ model.
The $Z_2$ critical exponents of the $XY$-Ising model on the critical line
have been compared with those of the frustrated $XY$ model on
square and triangular lattices using the same numerical methods
\cite{lkg91}. In particular, the estimate of the thermal exponent,
$\nu$, was obtained by finite-size scaling of Monte Carlo data
\cite{lkg91} which is insensitive to the estimate of the critical
temperature, $T_{\rm c}$. In most of the other numerical work, the exponent
$\nu$ is determined in an indirect way with several fitting parameters
while this evaluation involved only a one-parameter fit to the data.
The results, $\nu=0.83, \eta=0.28$ and $\nu=0.85, \eta=0.31$ for the
triangular and square lattice, respectively, are significantly
different from the pure Ising values and thus favor a single-transition
scenario. They are also in fair agreement with the critical exponents
in Table \ref{xyitab} for  the $XY$-Ising model on the critical line.
These were the first results which clearly indicated that the chiral
exponents deviate from pure Ising critical behavior. Independent
estimates using finite-size scaling of correlation functions
\cite{jose92}, Monte Carlo transfer matrix calculations \cite{gn93},
simulations of the lattice coulomb gas \cite{lee94} and exact transfer
matrix calculations  \cite{knops94} seems to agree on the estimate of
$\nu$ and the deviation of the chiral exponents from the pure Ising
behavior.

The deviation from the pure Ising exponents for the square lattice has
recently been questioned \cite{olsson} on the basis of a possible
failure of the finite-size scaling used in  previous estimates. Under
the assumption that the transition takes place in two stages very close
in temperature, another divergent length scale should be included in
the scaling analysis which could be the reason for the non-Ising
exponents reported so far. The result for the temperature dependence of
the $Z_2$ correlation length \cite{olsson} appear to be consistent with
the pure Ising exponent $\nu=1$ but we note that this calculation was
performed in a temperature range near $T_{\rm c}$ which is
significantly greater than the one where deviations from the Ising
exponent have been found \cite{gn93,knops94}. Due to the nearby Ising
and $XY$ transitions as in Fig. \ref{phadiag} and crossover effects, a
possible non-trivial fixed point on the single line can only be reached
if calculations are carried out close enough to the critical line which
requires both large systems and temperatures very close to $T_{\rm c}$.
Other calculations of the $Z_2$ correlation length but closer to
$T_{\rm c}$ are  consistent with deviations from the pure Ising behavior
\cite{jose92}.

Table \ref{ffxytab} summarizes recent results obtained for the
chiral order parameter of the frustrated $XY$ model.

\begin{table}[tbh]
\caption{Recent estimates for the chiral critical exponents $\nu$, $\eta$
and central charge $c$ of  the frustrated $XY$ model on a square and
triangular(T) lattice. Results considered consistent with $\nu=1$ are indicated
by $\sim 1$}
\medskip
\begin{center}
\begin{tabular}{c|ccc}
\hline \hline
\,                    & $\nu$             &$\eta$        &$c$        \\
\hline
Ref. \cite{tk90}       & $\sim 1 $         & $0.40$      &$1.66$     \\
Ref. \cite{lkg91}      & $0.82 $           & $0.31$      &\,         \\
Ref. \cite{lkg91}      & $0.83 $           & $0.28$  (T) &\,         \\
Ref. \cite{nico91}     & $\sim 1 $         & $0.26$      &\,         \\
Ref. \cite{jose92}     & $0.86 $           & $0.22$      &\,         \\
Ref. \cite{gn93}       & $0.80 $           & $0.38$      &$1.61$     \\
Ref. \cite{lee94}      & $0.84 $           & $0.26$      &\,         \\
Ref. \cite{knops94}    & $0.77 $           & $0.28$      &$1.55$     \\
Ref. \cite{olsson}     & $\sim 1 $         & \,          &\,            \\
\hline \hline
\end{tabular}
\end{center}
\label{ffxytab}
\end{table}

\section{Effects of disorder}

In real arrays, disorder is always present due to fabrication
processes but it can also be deliberately introduced in order to study
its effects \cite{lobb}. Two kinds of disorder may be present:
random plaquette areas and positional disorder \cite{gk89,kg88}. These
types of disorder are irrelevant in zero applied magnetic field but
become important and have interesting effects at higher fields. Randomness
of plaquette areas is realized when the superconducting grains have random
sizes but the magnetic field is sufficiently low to permit a partial
Meissner effect in individuals grain so that the fluxes can be
regarded as independent random variables. This random-flux type of
disorder has been argued to lead to the destruction of
superconductivity and to a finite  correlation length \cite{gk89},
$\xi \sim 1/H$, which depends on the field $H$, but vortex pinning
\cite{larkin,garland} by disorder is likely to dominate the
superconducting behavior.  If only weak positional disorder is present,
phase coherence is possible within a range of temperature $T^- < T <
T^+ $ for sufficiently low fields or disorder \cite{gk89,rsn}. For an
average value of  flux quanta per plaquette, $f_o$, the maximum field
or disorder for which quasi-long-range phase coherence is possible is
bounded by $f_o\Delta \le 1/\sqrt{32 \pi}$ where the probability
distribution of the grain positions is $P(\vec u) \propto \exp (-\vec
u^2/2 \Delta^2)$ and $\vec u$ is the displacement of a grain from
its average position. The analysis also shows  that the behavior of the
superconducting transition at $T^+$ has the same features as the
Kosterlitz-Thouless
transition but with a non-universal jump of the helicity modulus. The
disorder-induced reentrant phase transition which is predicted to occur
at the lower temperature, $T^-$, does not take into account vortex
pinning by disorder. Since the dynamics is expected to be very slow,
experimental or numerical study of the this transition is difficult.

Experiments on proximity-coupled arrays with deliberate positional
disorder \cite{lobb} and numerical simulations \cite{chakra,forrest}
for integer $f_o$ appear to be consistent with the existence of a
critical field but find no evidence for reentrance.  A recent
renormalization group treatment \cite{korsh} reaches the
conclusion that the  transition to the normal phase is not reentrant
but vanishes at the same critical disorder.

For $f_o= n + 1/2$, the ordered array has a $Z_2$ symmetry in addition
to the $U(1)$ symmetry. With a Coulomb-gas representation it has been
shown that random plaquette areas act like random fields on the $Z_2$
order parameter and positional disorder acts like random bonds
\cite{kg88,gk89}. Consequently, random plaquette areas should destroy chiral
order and phase coherence for any amount of disorder. This result seems
to agree with energy balance arguments for the chiral ground state
\cite{moore}.  Positional disorder, however, is expected to affect  the
chiral order less dramatically and have different effects on the $Z_2$
and $U(1)$ excitations. Increasing the amount of disorder should reduce
the $XY$ transition temperature to below that of the Ising if the $Z_2$
and $U(1)$ degrees of freedom are treated separately.  Eventually,
increasing disorder should lead to a double transition scenario where
phase coherence is lost within the chiral ordered phase where domain
walls and corner charges should have no effect. This leads to the same
critical field as for integer $f_o$.  In the experiments \cite{lobb} on
proximity-coupled arrays, the critical field $f_{\rm c}(q)$ appears to be
independent of $q$ where $f_o=p/q$ which is consistent with the
proposed scenario \cite{gk89} if it is realized for all rational values
of $f_o$. Monte Carlo simulations of the frustrated $XY$ model on a
square lattice finds evidence for the splitting into separated $XY$ and
Ising transitions \cite{nico91} as predicted but no evidence for
re-entrance \cite{ccs}.

\section{Quantum ladder}
A one-dimensional ladder of Josephson junctions, at zero temperature,
undergoes a superconductor to insulator transition as a function of
charging energy $E_{\rm c} $ due to capacitive effects.
The transition is strongly affected by the
magnetic field \cite{kardar,eg90} through the value of $f$. As for
two-dimensional arrays, the universality class of this transition is a
problem of great interest specially in relation to experiments
\cite{geerligs89,vdzant92} and theoretical predictions of universal
properties \cite{fisher90,gk90}.  At $f=1/2$, the effective Hamiltonian
\cite{eg92,eg90} describing fluctuations from a commensurate phase of pinned
vortices is expected to be in the same universality class of the
$XY$-Ising model in the parameter space $A=B$. The location of
the cut through the phase diagram in Fig. \ref{phadiag} depends on the
ratio $E_x/E_y$ between the interchain $E_x$ and intrachain $E_y$
Josephson couplings of the ladder. Since, the superconducting-to-insulator
transition at $T=0$  is to be identified with the loss of
phase coherence, this transition in the decoupled region is in the
Kosterlitz-Thouless universality class, but in the single transition
region it is possibly in a new universality class, the same as for a
two-dimensional classical array at finite temperature.

The critical behavior of the quantum ladder at $f=1/2$ has been studied
using a Monte Carlo transfer matrix applied to the path-integral
representation of the model and a finite-size scaling analysis
\cite{eg92}. In this formulation, the one-dimensional quantum model is
mapped into a two-dimensional classical model with an extra dimension,
the time direction. The parameter $\alpha = \sqrt{E_y/E_{\rm c}}$ plays the
role of an inverse temperature in the classical model, where $E_{\rm c}$ is
charging energy. Near the critical point $\alpha_{\rm c}$, chiral order is
destroyed by quantum fluctuations with an energy gap vanishing as
$|\alpha - \alpha_{\rm c}|^\nu$.  At the critical point, the correlation
function decay as a power with an exponent $\eta$. The results for the
critical coupling $\alpha_{\rm c}$ and critical exponents $\nu$  and $\eta$
for two different values of the ratio $E_x/E_y$ are indicated in
Table \ref{laddertab}.
For equal interchain and intrachain couplings, $E_x = E_y$, the results
for the critical exponents differ significantly from the pure
two-dimensional Ising model and are in fair agreement with the
corresponding values at the single line of the $XY$-Ising model in Table
\ref{xyitab}. For $E_x/E_y = 3$, the results are in reasonable agreement
with pure Ising values. In the phase diagram of the $XY$-Ising model,
this corresponds to a path in the decoupled region where $XY$ and Ising
transitions can take place at different points. In fact, for this case,
the values obtained for the critical coupling associated  with
chiral order $\alpha_I = 1.16 $, and the one associated with the
universal jump in the helicity modulus, $\alpha_{XY}=1.29$, clearly
indicate that there are two transitions. Somewhere in between $E_x/E_y
=1$ and $3$ there should be a bifurcation point where a single
transition with simultaneous loss of phase coherence and chiral order
decouples into separate transitions, the chiral transition being
located in the insulating phase.

\begin{table}[tbh]
\caption{Chiral critical exponents ($\nu, \eta $), for the quantum
ladder \protect \cite{eg92}. }
\medskip
\begin{center}
\begin{tabular}{c|cc}
\hline \hline
$E_x/E_y$            &$\nu$        &$\eta$    \\
\hline
1                    &$0.81(4)$    &$0.47(4)$  \\
3                    &$1.05(6)$    &$0.27(3)$  \\
\hline \hline
\end{tabular}
\end{center}
\label{laddertab}
\end{table}

\section{Conclusions}
Numerical results from Monte Carlo simulations and transfer matrix
computations support the identification of the critical behavior of
Josephson-junctions arrays at $f=1/2$ flux quantum per plaquette with
the universality class of the XY-Ising model.  For classical arrays
without charging effects, in the form of square and triangular
lattices, the results are consistent with a transition corresponding to
a cut through the XY-Ising phase diagram intersecting  the critical
line of simultaneous XY and Ising transitions. Some conflicting results
regarding the single or double nature of the transition may be
explained as resulting from the proximity of the cut to the branch
point in the phase diagram. For the quantum ladder, the ratio between
interchain and intrachain Josephson couplings can be used to tune the
transition near the branch point resulting in a simultaneous or double
transition depending on this ratio.  For classical arrays, weak random
plaquette disorder acts like a random field and positional disorder as
random bonds on the $Z_2$ order parameter. Positional disorder tend to
decouple the XY and  Ising variables leading to the same
critical disorder as for integer $f$.

\section{Acknowledgments}
This work was supported by the IAE Agency/ICTP (E.G.),  by a joint
NSF-CNPq Grant (E.G. and J.M.K.) and by the NSF Grants
No. DMR-92-22812 (J.M.K) and DMR-9214669 (M.P.N.).

\newpage

\end{document}